# Dielectric Properties of Composites Containing Melt-extracted Co-based Microwires

Yang Luo,[a] Faxiang Qin,[b (1)] Jingshun Liu,[c] Huan Wang,[b] Fabrizio Scarpa,[a] Pierre Adohi,[d] Christian Brosseau[d] and Hua-Xin Peng[b (2)]

[a]*Advanced Composite Centre for Innovation and Science, Department of Aerospace Engineering, University of Bristol, University Walk, Bristol, BS8 1TR, UK*
[b]*Institute for Composites Science Innovation (InCSI), School of Materials Science and Engineering, Zhejiang University, Hangzhou, 310027, China*
[c]*School of Materials Science and Engineering, Inner Mongolia University of Technology, China*
[d]*Lab-STICC, Université de Brest, 6 avenue Le Gorgeu, CS 93837, 29238 Brest Cedex 3, France*

## Abstract

We have investigated the microwave properties of epoxy-based composites containing melt-extracted $Co_{69.25}Fe_{4.25}B_{13.5-x}Si_{13}Nb_x$ ($x=0, 1, 3$) microwires of various length annealed using a so-called combined current-modulation annealing (CCMA) technique. The observation of a double-peak feature in the permittivity spectra is believed due to the coexistence of the amorphous phase and a small amount of nanocrystallites on the wires with a high Nb content. CCMA was found to be favorable for a better-defined circular anisotropy of microwires and had suppressed the high-frequency peak due to residual stress relief for the composite with 25 mm long wires. Neither the shift of resonance peak nor the characteristic double peak feature was detected for composites containing as-cast 15 or 35 mm long microwires.

---

[1]To whom the correspondence should be addressed.
 () faxiangqin@zju.edu.cn
[2] () hxpengwork@zju.edu.cn

## 1. Introduction

Amorphous ferromagnetic microwires have been extensively researched for sensing applications such as magnetic and stress sensors [1]. Amongst existing techniques for microwires production, microwires processed by melt-extraction (MET) technique possess superior mechanical and soft magnetic properties owing to the ultra-high cooling rate in the melt extraction process [2]. The quest for efficient tunable sensing and frequency-agile materials has led to the investigation of polymer-based composites containing such MET wires, where a variety of emerging functionalities have been reported [3].

Generally, there are two ways to tailor the electromagnetic (EM) performance of microwires. Firstly, doping of CoFe-based microwires using elements such as niobium is favorable for soft magnetic behavior [4]. Secondly, several post-annealing techniques have afforded marked control and tunable properties of the microwires in response to incident EM waves [5]. Nevertheless, conventional current annealing techniques have their own limitations. For example, direct current (DC) annealing generates excessive heat that may introduce damage to the wires; while the pulse current (PC) annealing fails to provide the persistent power required to optimize the domain structure. To overcome these issues, an optimized combination of these two annealing techniques, named the combined current-modulation annealing (CCMA), has been shown to provide a good compromise [6]. Understanding the influence of CCMA on the microwave properties of polymer-based composites containing microwires is rather challenging since the wire-polymer interface brings in extra complexity and the relations between wire length, annealing treatment and EM properties has been rarely revealed.

Against this background, as part of an initiative to the development of ferromagnetic microwire-composites for a range of crucial engineering applications such as structural health monitoring and microwave absorption, the present work aims to examine the microwave properties of polymer-based composites containing MET ferromagnetic Co-based microwires subjected to external magnetic field. The primary objective is to demonstrate the inter-dependences among chemical composition, CCMA, length of microwires and microwave response of microwire-composite with an emphasis on its dielectric permittivity.

## 2. Material and methods

Amorphous $Co_{69.25}Fe_{4.25}B_{13.5-x}Si_{13}Nb_x$ microwires (nominal values of $x$=0, 1, 3) with average diameter of 45 μm were synthesized via the MET technique. Details of the fabrication may be found in Ref. [2]. CCMA was performed by applying a PC at 50 Hz with amplitude of 90 mA for 480s, followed by a DC with amplitude of 65 mA passing through the wires for 480 s [6]. The wire samples were structurally characterized by HRTEM (JEM 2010F). To investigate the length effect, wires of 15, 25 and 35 mm were selected and randomly dispersed into epoxy (PRIME™ 20LV, Gurit UK) with a constant wire concentration of 0.026 vol. %, followed by a standard curing cycle [7]. For the microwave characterization, the composite samples have dimensions of $70×10×1.8$ mm$^3$ and are denoted as Nb0 ($x$=0), Nb1 ($x$=1), and Nb3 ($x$=3), respectively.

The effective complex (relative) permittivity $\varepsilon = \varepsilon'-j\varepsilon''$ was obtained from the measurement of the $S$ parameters in the frequency range from 0.3 to 6 GHz using a vector network analyzer (Agilent, model H8753ES). Instrumentation details can be found elsewhere [8]. An external magnetic bias was swept from 0 to ± 1.5 kOe by placing the transmission line between the poles of an electromagnet, which was monitored by using a Gauss meter equipped with a Hall element.

## 3. Results and discussion
### 3.1 Influence of Nb doping

Figure 1 shows a double-peak feature of the $\varepsilon$ spectra of the Nb3 sample containing as-cast or CCMA 25 mm microwires. The key to understanding the unique high frequency permittivity signature of the Nb3 sample is the microstructural change with Nb doping. Two arguments are in order here: Firstly, the inhomogeneously localized residual stresses initiate a nanocrystallite nucleation process; secondly, the growth of this nanocrystalline phase, which is strongly metastable, is inhibited by the rapid cooling rate during the fabrication process. Meanwhile, the amount of Nb is decisive to the nanocrystallite nucleation number. Nb is an efficient inhibitor for crystalline growth as it is rejected from the crystallization front due to its larger atom size and smaller diffusivity [4]. For doping content of 3%, the nucleation sites on the stress concentration locations initiate the nanocrystallization and the Nb atoms provide thermal stability to those formed nanocrystallites as demonstrated by previous studies [4]. The effective permittivity of such dual-phase structure can be described by $\varepsilon = \beta \varepsilon_{amor} + (1-\beta) \varepsilon_{nano}$, where $\beta$ is the volume fraction of the amorphous phase, $\varepsilon_{amor}$ and $\varepsilon_{nano}$ indicating the intrinsic permittivity of amorphous and nanocrystalline phases of microwires, respectively [9]. This hypothesis is validated by the HRTEM image of as-cast Nb3 microwires (Fig. 2(a)), which displays a small amount of nanocrystalline phase of ~2 nm in size. As also observed in Fig. 2(b) corresponding to the CCMA Nb3 sample, a larger amount of this nanocrystalline phase is formed and embedded in the amorphous phase together with a typical polycrystalline ring detected in the FFT pattern. The average size of the nanocrystalline phase stabilizes as 1.5 to 2 nm due to the inhibiting of crystalline growth arisen from the presence of niobium.

**3.2 Influence of CCMA**

Figure 3 displays the effective permittivity spectra of sample Nb1 containing 25 mm as-cast and CCMA microwires. A low-frequency peak is identified around 4 GHz in the composite containing as-cast microwires at zero field (Figs. 3(c) and 3(d)). When a magnetic bias is applied, a high-frequency peak is observed in the 5.0-5.5 GHz depending on the nominal field value. However, such a peak is suppressed in the specimen containing CCMA wires except for the highest values of the applied field. For brevity, the complex permittivity of Nb0 sample containing 25 mm CCMA wires is not shown here as it is very similar to the dielectric characteristics of Nb1 sample.

First of all, the observed low frequency peak is related to dipole resonance whose spectral position can be determined by $f = c/2l\sqrt{\varepsilon_m}$ Error: Reference source not found, where $c$, $l$ and $\varepsilon_m$ are respectively the light velocity in vacuum, wire length, and the permittivity of matrix [7]. Taking $l$=25 mm and $\varepsilon_m$ =2.5 [7], $f$ is calculated to be 3.8 GHz, which is close to the observed resonance peak. Now to address the high-frequency peak, it is important to note that such a peak is irrelevant to the nanocrystallites obtained in Nb3 sample due to the insufficient number of nucleation sites. Rather, the magnetic resonance would be the explanation. The present as-cast MET Co-based microwires are featured with a longitudinal anisotropy which is similar to that of glass-coated Fe-based wires. This is confirmed by a recent study that shows large ratio of the remanence to saturation magnetization ($M_s$) of MET Co-based wires ≈0.42 [10], implying a nearly longitudinal anisotropy that favors ferromagnetic resonance (FMR) [11]. The FMR frequency is then characterized asError: Reference source not found, where Error: Reference source not found and Error: Reference source not found (Error: Reference source not found<<$M_s$ for Fe-based wires) are the gyromagnetic ratio and anisotropy field, respectively [12]. However, as the external field is increased, this peak position does not shift to higher frequencies (Figs. 3(c) and 3(d)) as otherwise predicted by Kittel's relation [12]. This is owing to the residual stresses on the highly amorphous wires induced during the fabrication of wires, which manifests as additional contribution to the $H_a$ hence the $f_{FMR}$ [13]. CCMA allows a consistently thermal modification of domain structure and hence a stabilized Error: Reference source not found due to the relaxation of residual stresses. This explains why the FMR peak merges with the dipole resonant peak at low frequencies between 3 and 4 GHz after CCMA treatment (Fig. 3(d)).

### 3.3 Influence of wire length

Another crucial aspect meriting a detailed study is the mesostructure of the wire-composites, which we approach to from wire length as one of the important mesostrctural parameters. The transmission coefficients of the Nb1 composite containing different length of as-cast and CCMA microwires are compared in Fig. 4. For the composite containing 35 mm as-cast wires, we recall that the wire length is much larger than the sample width (10 mm). Consequently, the wires have formed an entangled state or even bundle during the curing cycle, which has been demonstrated in a previous study [14], yielding the dipole model invalid. This is consistent with the observation of significant enhancement of conductivity (Fig. 4(a)) due to strong magnetostatic interactions among wires and their high magnetoelastic energy [15]. The reason behind the improved field-dependence of permittivity in Fig. 4(b) is that CCMA has a positive effect of significantly reducing the wire-wire magnetoelastic interactions in composite containing 35 mm long wires. However, the influence of CCMA is not the single factor here because no blueshift of the resonance peaks is found in the transmission spectrum of the specimen filled with CCMA wires. A possible way to avoid wire entanglement is to select shorter wires. Nevertheless, the complex nature of the $S_{21}$ spectrum for the composite containing as-cast 15 mm wires (Fig. 4(c)) precludes a precise explanation. Imperfect interfacial bonding between the microwires and epoxy matrix can be a possible cause for this trend. If now CCMA is performed, the residual stresses in the as-cast wires are released, leading to the restoration of the dielectric features (Fig. 4(d)).

Based on the experimental results, a chemical treatment of the microwires after CCMA to modify the internal stresses and further to improve the wire/epoxy interface bonding would be a major avenue for future work. A treatment of microwires by Si-based chemical agents such as silane, for instance, is likely to improve their soft magnetic properties and interfacial conditions with epoxy by creating strong covalent Fe-O-Si bonding [16, 17].

## 4. Conclusions

In summary, we have quantified the dielectric properties of composites containing MET ferromagnetic microwires. A double-peak feature is noted in the permittivity spectrum of Nb3 sample due to the presence of amorphous-nanocrystalline microstructure of the wires. CCMA proved to be an effective approach to release internal stresses in the wires and make the dielectric behavior of wire-composites more predictable. In the present work, 25 mm manifests to be the optimal wire length whereas further chemical treatment by silane would improve the dielectric properties of composites containing long or short wires. These findings provide a promising route towards direct magnetic field control of permittivity at the micro- and meso-scale, underscoring the effective utilization of emergent electromagnetic properties in complex heterostructural composites.

**Acknowledgements**

Y. Luo acknowledges financial support from University of Bristol Postgraduate Scholarship and China Scholarship Council. FXQ would like to thank the support from NSFC under grant No. 51501162. Lab-STICC is UMR CNRS 6285.

tunable microwave properties of composites containing Fe-based microwires. Appl. Phys. Lett. 104 (2014) 121912.

**Figure captions**

Fig. 1 (color online) Frequency plots of the $\varepsilon$' of Nb3 sample containing 25 mm (a) as-cast and (b) CCMA wires, and the $\varepsilon$" of Nb3 sample containing 25 mm (c) as-cast and (d) CCMA wires.

Fig. 2 (color online) HRTEM images of (a) as-cast and (b) CCMA Nb3 microwires. The insets represent the FFT patterns of selected areas of microwires.

Fig. 3 (color online) (a) Spectra of the real part of the effective permittivity ($\varepsilon$') of Nb1 specimen containing 25 mm as-cast microwires as a function of applied magnetic field; (b) As in (a) for CCMA microwires; (c) Spectra of the imaginary part of effective permittivity ($\varepsilon$") of Nb1 specimen containing 25 mm as-cast microwires as a function of applied magnetic field; (d) As in (c) for CCMA microwires.

Fig. 4 (color online) Transmission coefficients, $S_{21}$, of Nb1 sample containing 35 mm (a) as-cast and (b) CCMA wires and 15 mm (c) as-cast and (d) CCMA wires.

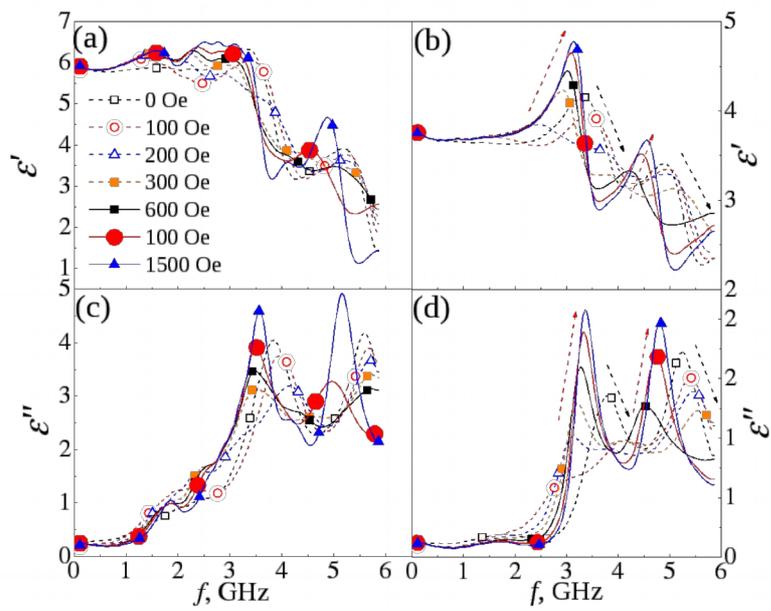

FIG1

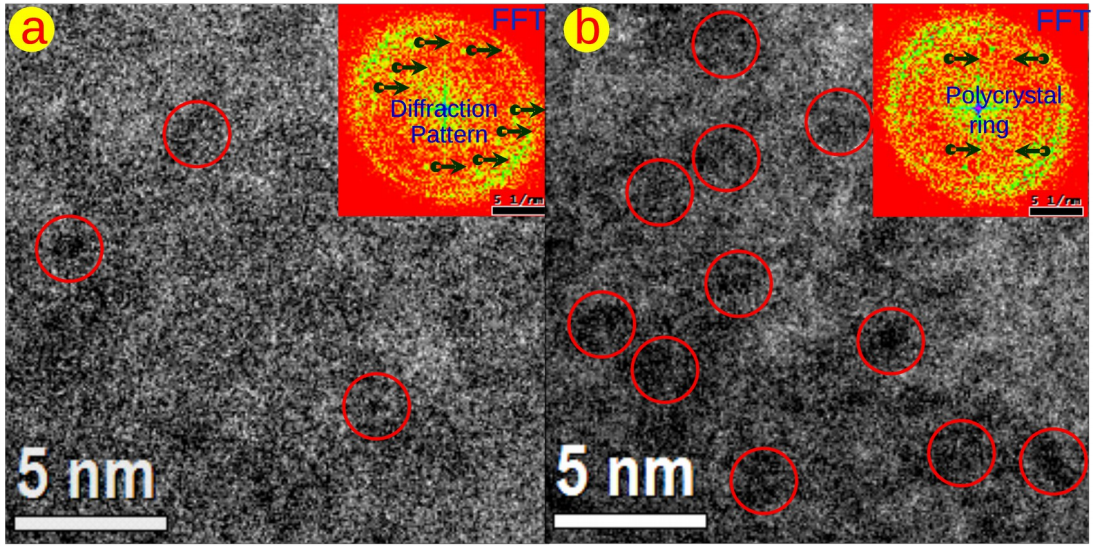

FIG2

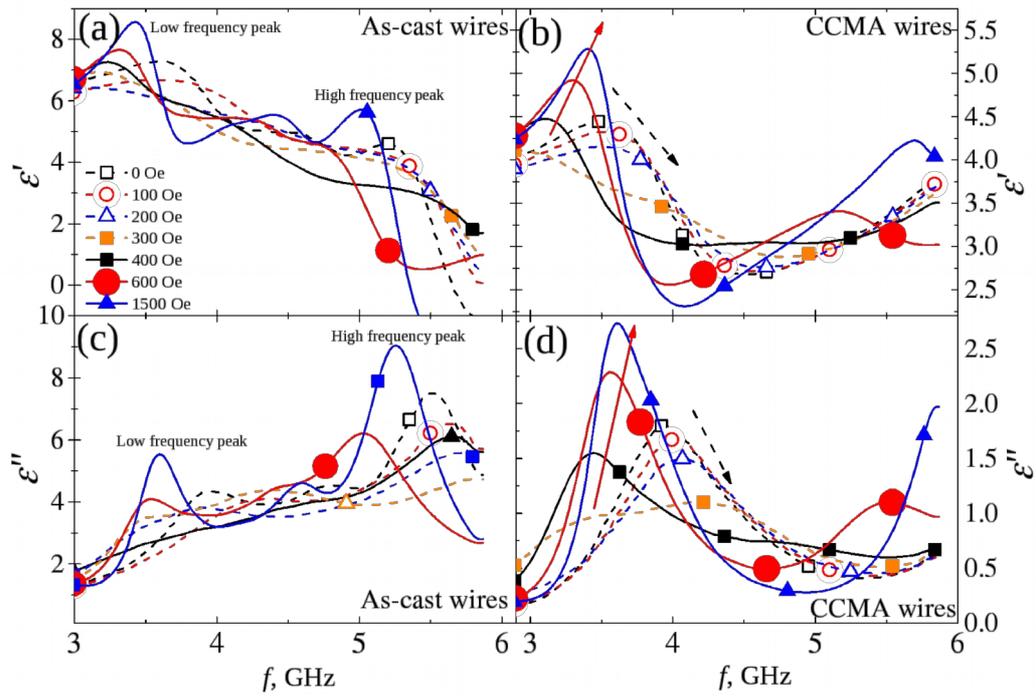

FIG3

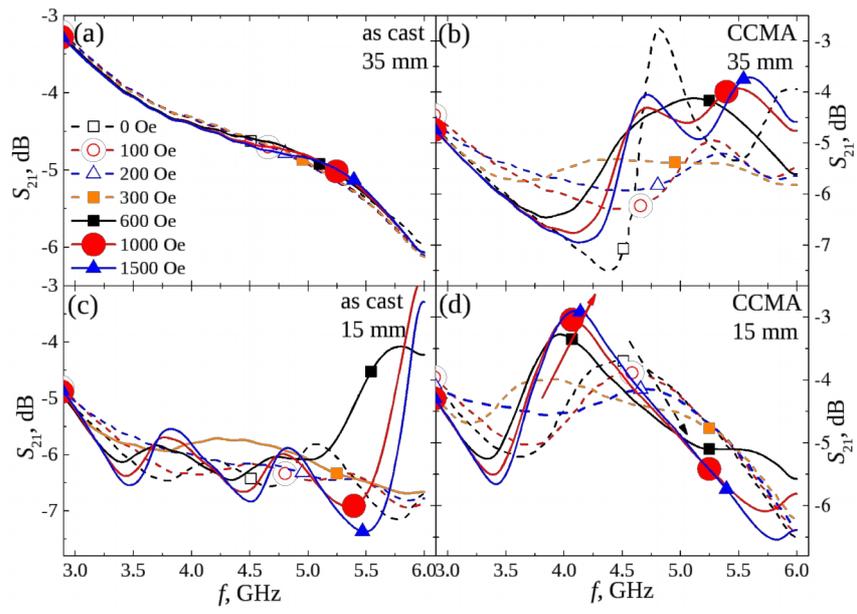

FIG4